\documentclass[iop,apj,12pt]{aastex63}
\usepackage{apjfonts}
\usepackage{lineno}
\usepackage[normalem]{ulem}
\usepackage{verbatim}
\usepackage{ulem}
\usepackage{graphicx}
\usepackage{hyperref}
\usepackage{soul}
\usepackage{gensymb}

\shortauthors{CROMARTIE ET AL.}
\shorttitle{Shapiro Delay Measurements of PSR \psr{}}

\begin{document}

\def\fermi{{\em Fermi}}
\def\msun{M$_{\odot}$}
\def\psr{J1231$-$1411}
\def\mc{$m_{\rm c}$}
\def\mp{$m_{\rm p}$}
\def\sini{sin($i$)}

\newcommand{\htc}[1]{\textcolor{red}{\bf{\textbf{#1}}}}

\title{Shapiro Delay Measurements from Fifteen Years of PSR \psr{} Radio Observations}

\correspondingauthor{H.~Thankful Cromartie}
\email{thankful.cromartie@nanograv.org}

\author[0000-0002-6039-692X]{H.~Thankful Cromartie}
\affiliation{National Research Council Research Associate, National Academy of Sciences, Washington, DC 20001, USA; resident at Naval Research Laboratory, Washington, DC 20375, USA}

\author[0000-0002-0893-4073]{Matthew Kerr}
\affiliation{Space Science Division, Naval Research Laboratory, Washington, DC 20375-5352, USA}

\author[0000-0001-5799-9714]{Scott M.~Ransom}
\affiliation{National Radio Astronomy Observatory, 520 Edgemont Road, Charlottesville, VA 22903, USA}

\author[0000-0002-5297-5278]{Paul S.~Ray}
\affiliation{Space Science Division, Naval Research Laboratory, Washington, DC 20375-5352, USA}

\author[0000-0002-9049-8716]{Lucas Guillemot}
\affiliation{LPC2E, OSUC, Univ Orleans, CNRS, CNES, Observatoire de Paris, F-45071 Orleans, France}
\affiliation{ORN, Observatoire de Paris, Universit\'e PSL, Univ Orl\'eans, CNRS, 18330 Nan\c{c}ay, France}

\author[0000-0002-1775-9692]{Isma\"{e}l Cognard}
\affiliation{LPC2E, OSUC, Univ Orleans, CNRS, CNES, Observatoire de Paris, F-45071 Orleans, France}
\affiliation{ORN, Observatoire de Paris, Universit\'e PSL, Univ Orl\'eans, CNRS, 18330 Nan\c{c}ay, France}

\author[0000-0001-8384-5049]{Emmanuel Fonseca}
\affiliation{Department of Physics and Astronomy, West Virginia University, P.O. Box 6315, Morgantown, WV 26506, USA}

\author[0000-0002-3649-276X]{Gilles Theureau}
\affiliation{LPC2E, OSUC, Univ Orleans, CNRS, CNES, Observatoire de Paris, F-45071 Orleans, France}
\affiliation{ORN, Observatoire de Paris, Universit\'e PSL, Univ Orl\'eans, CNRS, 18330 Nan\c{c}ay, France}

\begin{abstract}

We present 15 years of Nan\c{c}ay and Green Bank radio telescope timing observations for PSR \psr{}. This millisecond pulsar is a primary science target for the Neutron Star Interior Composition Explorer telescope (NICER, which discovered its X-ray pulsations), has accumulated near-continuous $\gamma$-ray data since the \fermi-Large Area Telescope's launch, and has been studied extensively with the Green Bank and Nan\c{c}ay radio telescopes. We have undertaken a campaign with the Green Bank Telescope targeting specific orbital phases designed to improve our constraint on the pulsar's mass through the detection of a relativistic Shapiro delay. 
Both frequentist and Bayesian techniques---the latter incorporating priors from white dwarf binary evolution models---are applied to fifteen years of radio observations, yielding relatively weak constraints on the companion and pulsar masses of $0.23^{+0.09}_{-0.06}$ \msun{} and $1.87^{+1.11}_{-0.67}$ \msun{}, respectively (68.3\% CI from Bayesian fits); however, the orbital inclination is measured to better relative precision ($79.80^{+3.47}_{-4.70}$\degree). Restricting the maximum allowed pulsar mass to 3\,\msun{} improves the constraint and lowers the measured mass to $1.71^{+0.70}_{-0.56}$ \msun{}. A fully-generalized Bayesian fit that simultaneously samples the noise and timing models yields a pulsar mass in close agreement with this value. While our radio-derived inclination result has informed recent NICER X-ray studies of PSR \psr{}, the lessons learned from this troublesome pulsar will also bolster future high-precision mass measurement campaigns and resulting constraints on the neutron star interior equation of state.

\end{abstract}


\section{Introduction} \label{sec:intro}

Multi-messenger observations in the electromagnetic and gravitational wave bands play a critical role in the effort to determine the equation of state (EoS) of supranuclear-density matter. Observations of relativistic Shapiro delay in radio millisecond pulsars (MSPs) have resulted in precise measurements of the heaviest neutron stars' masses \citep[e.g.][]{dem10,fon21}, imposing a direct constraint on the neutron star interior EoS. 

X-ray light curve modeling, facilitated by instruments such as NASA's Neutron Star Interior Composition Explorer (NICER) telescope \citep[e.g.][]{ril19,mil19,gen17}, can constrain the EoS through the simultaneous measurement of masses ($M$) and radii ($R$).  
General relativistic effects from strong gravity affect the X-ray emission from hot spots on the neutron star surface in ways that depend strongly on the stellar compactness.  The substantial bending permits viewers to see over the neutron star horizon \citep{bog07}.
The precise modeling of the observed light curve therefore permits extracting the mass-to-radius ($M/R$) ratio for the neutron star \citep[$M$ and $R$ can be individually determined to a certain extent, but an independent measure of $M$ improves the estimation of these parameters; an overview is presented in][]{wat16}. 

One such study by \cite{raj21} combines neutron star mass-radius constraints from NICER and the XMM-Newton European Photon Imaging Camera \citep{ril21}, radio Shapiro delay observations \citep{cro20, fon21}, compact binary mergers GW170817 and GW190425 \citep{abb19a, abb20a}, and the kilonova AT2017gfo \citep{kas17}. In constraining the mass-to-radius ratio, they find that a 1.4-\msun{} neutron star will have a radius of $12.33^{+0.76}_{-0.81}$\,km or $12.18^{+0.56}_{-0.79}$\,km for piecewise-polytropic and speed-of-sound EoS parameterizations, respectively. Such an analysis is a testament to the power of a multi-messenger observational approach to constraining the neutron star interior EoS.

We present a pulsar timing analysis from radio observations of the MSP \psr{}, finding that a significant measurement of Shapiro delay in the radio timing data constrains the pulsar's mass, albeit to poor precision given the orbital geometry and timing properties of this system. The measurements reported in this manuscript served as conservative priors in NICER's light curve modeling for this source, which helps break the degeneracy in the \emph{M/R} ratio and allows discrimination among competing models of hot spot geometry \citep{sal24}. Section~\ref{sec:obs} provides an overview of the Green Bank Telescope (GBT) and Nan\c{c}ay Radio Telescope (NRT) observations, and describes available data from the \emph{Fermi}-Large Area Telescope (LAT). Section~\ref{sec:methods} discusses the derivation of pulsar masses via the relativistic Shapiro delay as well as software used in this analysis. Section~\ref{sec:analysis} presents the noise modeling and timing analysis of the data, as well as the various Bayesian approaches taken in deriving the mass of PSR \psr{}. 
A discussion (Section~\ref{sec:discussion}) and conclusions (Section~\ref{sec:conclusion}) follow.

\section{Observations \& Initial Analysis}\label{sec:obs}

\begin{figure}
    \centering
    \includegraphics[width=0.8\textwidth]{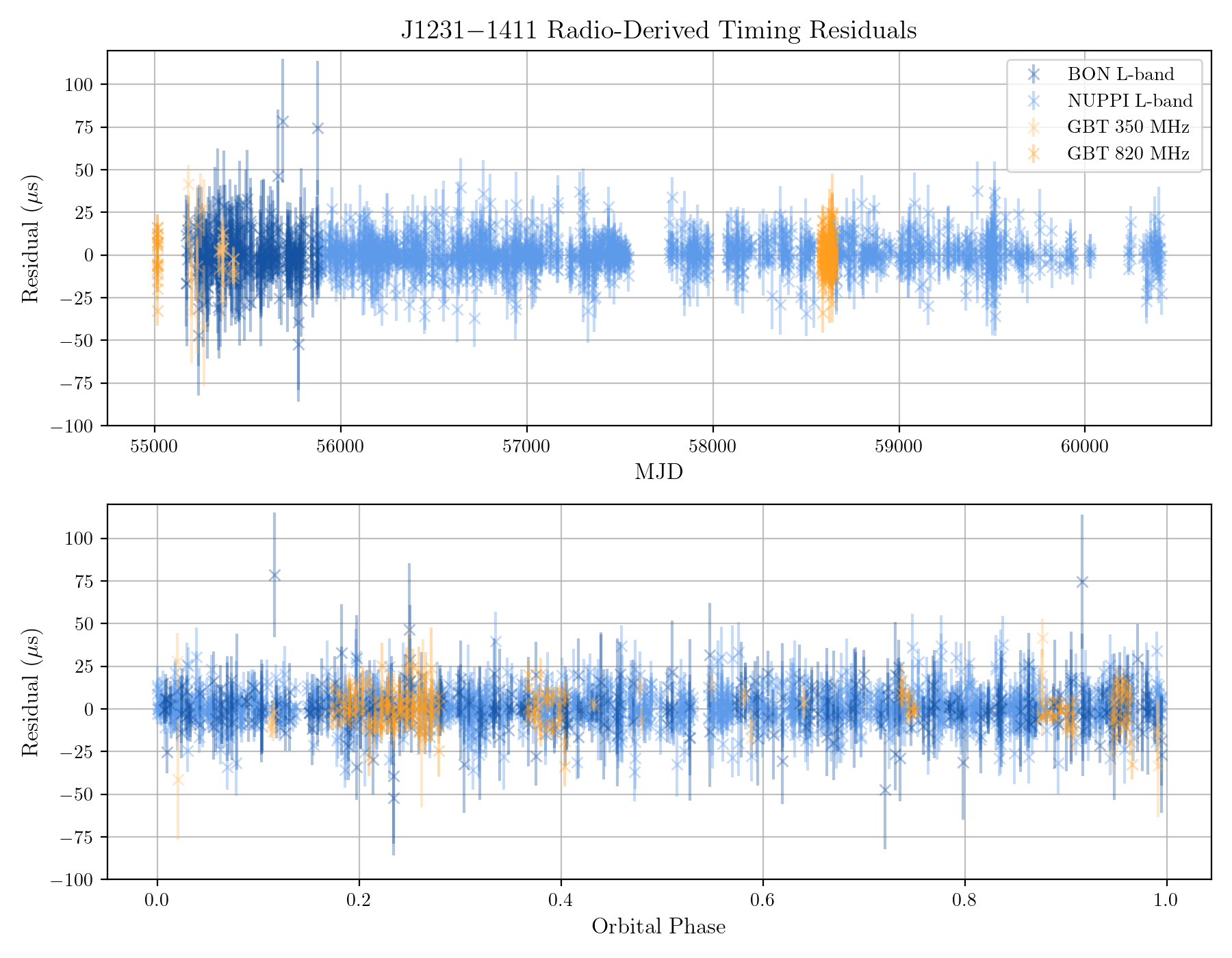}
    \caption{Timing residuals in $\mu$s vs.~MJD (top panel) and Orbital Phase (bottom panel) for \psr{} radio data. All best-fit timing parameters, including those that describe the relativistic Shapiro delay, are included in the model that produces these residuals. Color groups reflect backend and receiver combinations. Dark blue points are from NRT's BON backend and light blue points are from NRT's NUPPI backend; both of these groups use the same L-band receiver and are integrated into 2 and 4 sub-bands, respectively. Darker and lighter orange points are sub-banded GBT 820-MHz TOAs and 350-MHz TOAs, respectively, both of which used the GUPPI backend.}
    \label{fig:resids}
\end{figure}

\psr{} was discovered with the GBT during a radio pulsar search campaign targeting 25 of the brightest unassociated \emph{Fermi}-LAT $\gamma$-ray sources\footnote{In the \emph{Fermi}-LAT Bright Source List; see \cite{abd09}.} \citep{ran11}. This pulsar is a relatively typical 3.68-ms MSP in a 1.86-day orbit with a white dwarf companion (projected semi-major axis of the pulsar's orbit = 2.043\,lt-s). Since 2011, \psr{} has been intensively observed at radio wavelengths with the NRT, and more sporadically, the GBT (see Figure~\ref{fig:resids}). Although the timing precision of this source is better at 820 MHz than at L-band (1.4\,GHz), strong diffractive scintillation sometimes renders this pulsar invisible at lower frequencies. \cite{ray19} reported X-ray pulsations from \psr{}, with initial spectral modeling suggesting thermal pulsations originating from at least one hot spot on the neutron star's surface.

\subsection{Green Bank Telescope}
After its discovery at 820 MHz with the GBT in July 2009, a timing campaign for PSR \psr{} was conducted between December 2009 and August 2010 using the GBT 350-MHz receiver \citep{ran11}. Observing epochs were frequent and evenly spaced throughout the entire timing campaign.

We present a new set of observations from the GBT taken in Spring 2019 (project code GBT19A-429) and analyzed in combination with $\sim$15 years of additional radio timing data. The 22-hour GBT project comprised five, two-hour observations scheduled at random orbital phases, and two, six-hour observations over superior conjunction (where the Shapiro delay is most prominent). These observations were taken with the 820-MHz receiver and the Green Bank Ultimate Pulsar Processing Instrument backend \citep[GUPPI;][]{dup08} with 200 MHz of bandwidth split into 512 channels, later integrated into 128-channel groups, for a total of 4 sub-bands. All observations were conducted using an online folding mode with coherent dedispersion. Portions of several observations were removed, as diffractive scintillation sometimes rendered the MSP too dim. Notably, the superior conjunction observations saw the MSP scintillated down, reducing the benefit of the orbital-phase-targeted campaign. Calibrated total-intensity data were integrated over $\sim$16-minute (1000\,s) intervals, and the resulting profiles were matched to a smoothed template using \texttt{PSRCHIVE} \citep{van11} to obtain time-of-arrival (TOA) measurements (see \citeauthor{arz18}~\citeyear{arz18} for a detailed explanation of this process). The median TOA uncertainty for all GBT data is 6.2\,$\mu$s.

\subsection{Nan\c{c}ay Radio Telescope}
PSR \psr{} has been observed near-continuously with the NRT since November 2009, with the Berkeley-Orl\'{e}ans-Nan\c{c}ay (BON) backend until August 2011, and with the Nan\c{c}ay Ultimate Pulsar Processing Instrument (NUPPI) backend after it became the primary pulsar timing instrument in August 2011 \citep{cog13, des11}. BON observations were coherently dedispersed and carried out over a bandwidth of 128~MHz. NUPPI observations were also coherently dedispersed and conducted over an increased bandwidth of 512~MHz. The bulk of BON (respectively, NUPPI) observations were conducted at a central frequency of 1398~MHz (1484~MHz). Additional details on the NRT and on pulsar observations with this telescope can be found in \citeauthor{gui23} (\citeyear{gui23}). 

NRT observations were integrated in time, and the resulting profiles were then split into multiple frequency sub-bands. BON observations were integrated into two channels of 64~MHz each, and NUPPI observations were integrated into four channels of 128~MHz. Eight observations with the highest signal-to-noise ratios were then smoothed to produce standard profiles for each of the two NRT datasets. The polarization information was preserved throughout these data preparation steps. TOAs were finally extracted from the NRT observations using the Matrix Template Matching (MTM) method of \texttt{PSRCHIVE} \citep{van06}, which uses information in all four Stokes parameters to extract TOAs while compensating for differences between the profile and the template. 

While BON observations were directly time-tagged using a GPS-disciplined 1PPS signal (Trimble Thunderbolt), early NUPPI observations were compared to a freely-running 1PPS signal provided by a local Rubidium clock. Regular time comparisons with a GTR-50 receiver were conducted until MJD $\sim$58000 (September 2017), after which the Rubidium clock was also GPS-disciplined. A time drift beginning in 2014 was later found in the data produced by the GTR-50 receiver at Nan\c{c}ay. Nan\c{c}ay clock corrections that properly handled the GTR-50 receiver were used for this work. 

In addition, as documented in \citeauthor{gui23} (\citeyear{gui23}), the replacement of a pre-amplifier in 2019 significantly modified the polarimetric response of the NRT, which motivated a revision of the polarization calibration procedure in late 2019. Consequently, all NUPPI observations recorded after mid-2019 were calibrated using the significantly improved calibration procedure, while older NUPPI observations were calibrated with a more generic method. For the purposes of fitting and noise modeling, we therefore divided the NUPPI TOA datasets into two groups: pre- and post-MJD 58600 data, with a JUMP (time-offset parameter) between them. Nevertheless, the MTM method was shown to be effective at reducing timing noise resulting from imperfect polarization calibration \citep{gui23}.

\subsection{\emph{Fermi} Large Area Telescope}
The Large Area Telescope (LAT) is a GeV $\gamma$-ray instrument aboard NASA's \emph{Fermi} satellite, which was launched in June 2008 \citep{atw09}. PSR \psr{} is one of the two brightest $\gamma$-ray MSPs as seen from Earth \citep{abd13}, and is the best-timed MSP in \citeauthor{ker22} \citeyear{ker22}, the first data release of the ``Gamma Pulsar Timing Array'' (GPTA).  However, these data are most useful for studying long-timescale phenomena, whereas the radio data sets used in this work are ideal for the study of binary Shapiro delays. For comparison, when $\gamma$-ray data were integrated to produce radio-like TOAs, only two TOAs/year of comparable timing precision (to the thousands of TOAs used in our radio analysis) could be constructed from LAT data. An analysis of \emph{Fermi}-LAT data for PSR \psr{} will be included in the forthcoming second data release of the GPTA.

\section{Methods Overview}\label{sec:methods}
Among the many high-precision astrophysical measurements facilitated by pulsar timing is Shapiro delay, a general relativistic effect that is observable in a small subset of highly inclined MSP binaries \citep{sha64}. When an MSP passes behind its stellar companion along an observer's line of sight, the companion's spacetime deformation causes pulsations from the MSP to arrive later than would be otherwise expected. For circular orbits, the maximum delay occurs at orbital phase = 0.25 (superior conjunction), and is observable as a spike in timing residuals of order $\sim$10\,$\mu$s. By measuring the extent and shape of this delay, one obtains the post-Keplerian parameters range ($r$) and shape ($s$), the definition of which is dependent on one's assumed gravitational model. Here, we assume the validity of general relativity; see e.g.~\citeauthor{tay89}~\citeyear{tay89}; \citeauthor{dam92}~\citeyear{dam92}. When combined with the Keplerian mass function (Equation~\ref{eq1}), measurements of $r$ and $s$ (Equations~\ref{eq2} and~\ref{eq3}, \citeauthor{tay92} \citeyear{tay92}) uniquely constrain the pulsar and companion masses \mp{} and \mc{}: 
\begin{equation}\label{eq1}
f(m_{\rm p}, m_{\rm c}) = \frac{4\pi^2}{G} \frac{(a_{\rm p}\,{\rm sin} i)^3}{P_{\rm b}^2} = \frac{4\pi^2}{T_{\odot}}\frac{x^3}{P_{\rm b}^2} = \frac{(m_{\rm c} \,{\rm sin}i)^3}{(m_{\rm p} + m_{\rm c})^2} 
\end{equation}
\begin{equation}\label{eq2}
r = T_{\odot} m_{\rm c}
\end{equation}
\begin{equation}\label{eq3}
s = x\left(\frac{P_{\rm b}}{2\pi}\right)^{-2/3}T_{\odot}^{-1/3}(m_{\rm p} + m_{\rm c})^{2/3}m_{\rm c}^{-1} = {\rm sin}(i)
\end{equation}
where $T_{\odot}$ = $GM_{\odot}/c^3$ = 4.925490947\,$\mu$s (to yield a result in solar masses), $P_{\rm b}$ is the binary orbital period, and $x \equiv a_{\rm p}\sin i/c$.
Although pulsar timing very precisely determines the mass function, the precise measurement of post-Keplerian parameters is one of the most reliable ways to determine the companion and pulsar masses independently.
Early data from the GBT and NRT permitted an initial measurement of Shapiro delay that suggested the system was highly inclined, and that the MSP had a relatively low mass. Though EoS constraints from neutron star mass measurements \emph{alone} require the observation of more and more massive MSPs, the precise measurement of radii at a variety of neutron star masses constrains the neutron star EoS. Therefore, NICER measurements of $R$ for low-mass and high-mass MSPs alike can prove valuable.

This work employs a variety of standard pulsar timing tools to constrain the mass of \psr{}. For both Nan\c{c}ay and GBT observations, we use the \texttt{PINT} \citep{luo21} pulsar timing software package to fit sub-banded TOAs. \texttt{PINT} features both traditional $\chi^2$ fitting routines for pulse times of arrival (TOAs) and Markov Chain Monte Carlo (MCMC) routines. The specifics of these techniques are described in Section~\ref{sec:analysis}. Bayesian noise estimation of red and white noise parameters (see Section~\ref{sec:noise}) was performed using \texttt{ENTERPRISE} \citep{enterprise}. All fits employed the DE440 planetary ephemeris provided by NASA's Jet Propulsion Laboratory, and used the TDB timescale. 

\section{Analysis \& Results}\label{sec:analysis}

The data set presented in this work requires careful treatment for a number of reasons. 
Among these are the aforementioned strong scintillation, heterogeneous composition of backend and receiver combinations, and poor timing precision with significant white (achromatic, stochastic) noise. Covariances also exist between slowly-varying astrometric and general relativistic parameters and red (low-frequency) noise processes, such as DM variations (chromatic red noise) and borderline-significant intrinsic red noise (achromatic red noise; see Section~\ref{sec:noise}). 
Because the Shapiro delay's timing signature is strongly dependent on the orbital inclination angle, an 80$\,\degree$ tilt yields a notably weak Shapiro effect. If PSR \psr{} were instead inclined at $i$ = 90$\degree$, the measurable Shapiro delay would be $\sim$3 times larger and far less susceptible to timing and noise model changes.

Given the aforementioned complexities, a simple linear regression with Gaussian errors is not sufficient to understand how well constrained \mc{} might be in our \psr{} data set. Bayesian inference provides more thorough estimates of parameter uncertainties, especially in cases where the parameters of interest are highly susceptible to covariances, or have strongly non-Gaussian probability distributions. We employ three separate methods in this work:
\begin{itemize}
    \item A $\chi^2$ gridding scheme that performs a least-squares fit with Shapiro delay parameters fixed at each of a set of \mc{}-\sini{} pairs and with white and red noise parameter values measured \emph{a priori} (Section~\ref{sec:grid});
    \item A Bayesian fit with astrophysically-derived priors where all parameters, including the Shapiro delay parameters, are fit simultaneously but noise is measured \emph{a priori} (Section~\ref{sec:pintmcmc});
    \item An exploratory fully-generalized fit where the noise and timing models are determined simultaneously, with some timing parameters being treated as non-linear (Section~\ref{sec:nonlinearresults}).
\end{itemize}

In the first and second cases, the white noise parameters are determined with \texttt{ENTERPRISE} prior to the timing model fit(s). The third case uses \texttt{ENTERPRISE}, which calls upon \texttt{PINT}, to simultaneously sample the timing and noise models. 

The ``Measured Value'' column of Table~\ref{tab:1231_params} presents the results of a typical \texttt{PINT} linear regression fit of the combined data set, including the Shapiro delay parameters \mc{} and \sini{}. We employed the binary timing model ELL1 \citep{lan01} for this nearly circular MSP system. ELL1 includes the following orbital parameters:~projected semi-major axis ($x$); binary orbital period ($P_{\mathrm b}$); epoch of ascending node (TASC); and the two Laplace-Lagrange parameters, $\epsilon_1$ = $e$\,sin\,($\omega$) and $\epsilon_2$ = $e$\,cos\,($\omega$), the orbital eccentricity multiplied by the sine and cosine of periastron longitude, respectively. Because these values are already known to high precision, the greatest source of uncertainty in our analyses is measuring the more poorly constrained Shapiro delay parameters $m_{\mathrm c}$ and \sini{}, as well as astrometric (parallax, proper motion) and other post-Keplerian parameters ($\dot{P_{\rm b}}$). 

\subsection{Noise Analysis for Radio Data}\label{sec:noise}
Pulsar timing precision is additionally limited by stochastic white noise sources beyond radiometer noise, such as pulse jitter and radio frequency interference. We contend with these noise sources by including two white noise parameters in our fits for each observing frequency and backend combination included in our analysis (see Table~\ref{tab:WN}). The first is EFAC, a scaling error applied to TOA uncertainties; the second is EQUAD, a term added in quadrature to the TOA uncertainties. More details about the use of these parameters can be found in, e.g., \cite{arz18}. We note that ECORR, a third commonly used white noise parameter that accounts for inter-channel correlations among TOAs, is not employed here due to a lack of frequency coverage and resolution \citep{lam17}.

The white noise parameter estimations were derived using the \texttt{ENTERPRISE} modeling suite and frequency-resolved TOAs, with separate EFAC and EQUAD values measured for each backend-receiver combination. Early GBT 820-MHz observations are treated separately from the 2019 observations, and the NUPPI TOAs are split into two groups based on the significant instrumental changes in 2019. Measured white noise parameter values are presented in Table~\ref{tab:WN}. JUMP parameters are included for all backend-receiver combinations except the pre-58600 subset of NUPPI L-band data, relative to which all other JUMPs are measured. GBT and NRT data have relatively large and small associated offsets, respectively (with the exception of the poorly-constrained JUMP associated with the short 2009 GBT data set). This is expected when choosing the NRT data as the reference. 

In a subset of well timed MSPs, achromatic red noise (spin noise), likely caused by irregularities in neutron star rotation, can be significantly measured over long time scales. The power spectral density of this red noise is typically modeled with a power law, which has been shown to be sufficient for MSPs \citep{cord16, gon20}. A free-spectral model specified over 14 frequency bins was used for exploratory purposes during this work in an effort to pinpoint the origin of relevant noise processes \citep{len16, haz20}; however, the results presented here employed a more standard power-law red noise model with 30 frequency bins. Though each run tested for the presence of red noise, none met the inclusion criterion (Bayes factor >100) when $\dot{P_{\rm b}}$ and parallax were present in the timing model. Section~\ref{sec:discussion} includes a discussion of the impact of red noise model selection on MSP mass measurements. 

\begin{table}
\centering{
\caption{\label{tab:1231_params}PSR \psr{} Best-Fit Parameters from Radio Linear Regression}
\begin{tabular}{ll}
\hline

\hline
&  Measured Value \\
\hline
\multicolumn{2}{c}{Measured Quantities} \\ 
\hline
Pulsar name                  \dotfill & \psr{}\\ 
MJD range                    \dotfill & 55015---60408 \\ 
Data span (yr)               \dotfill & 14.77 \\ 
PX, Parallax ($\mathrm{mas}$)\dotfill &  $1.7(3)$ \\ 
RAJ, Right ascension (J2000) ($\mathrm{{}^{h}}$)\dotfill &  $12.519801168(4)$ \\ 
DECJ, Declination (J2000) ($\mathrm{{}^{\circ}}$)\dotfill &  $-14.1954434(1)$ \\ 
PMRA, Proper motion in RA ($\mathrm{mas\,yr^{-1}}$)\dotfill &  $-61.63(4)$ \\ 
PMDEC, Proper motion in DEC ($\mathrm{mas\,yr^{-1}}$)\dotfill &  $6.90(9)$ \\ 
F0, Spin-frequency ($\mathrm{Hz}$)\dotfill &  $271.453019265084(2)$ \\ 
F1, Spin-frequency derivative 1 ($\mathrm{Hz\,s^{-1}}$)\dotfill &  $-1.66710(3) \times 10^{-15}$\\ 
DM, Dispersion measure ($\mathrm{pc\,cm^{-3}}$)\dotfill &  $8.0845(4)$\\ 
DM1, first time derivative of DM ($\mathrm{pc\,yr^{-1}\,cm^{-3}}$)\dotfill &  $0.0005(1)$\\ 
DM2, second time derivative of DM ($\mathrm{pc\,yr^{-2}\,cm^{-3}}$)\dotfill &  $-0.00011(5)$\\ 
PB, Orbital period ($\mathrm{d}$)\dotfill &  $1.8601438836(1)$\\ 
PBDOT, Orbital period derivative respect to time\dotfill &  $7.6(5) \times 10^{-13}$ \\ 
A1, Projected semi-major axis of orbit ($\mathrm{ls}$)\dotfill &  $2.0426225(8)$ \\ 
M2, Companion mass (\msun{})\dotfill & $0.20(6)^a$\\ 
SINI, Sine of inclination angle\dotfill &  $0.989(9)^a$\\ 
TASC, Epoch of ascending node ($\mathrm{d}$)\dotfill &  $55015.15346563(9)$\\ 
EPS1, First Laplace-Lagrange parameter, ECC*sin(OM)\dotfill &  $-2.8(3) \times 10^{-6}$\\ 
EPS2, Second Laplace-Lagrange parameter, ECC*cos(OM)\dotfill &  $9(2) \times 10^{-7}$ \\ 
\hline
\multicolumn{2}{c}{Set Quantities} \\ 
\hline
Reference epochs for position, spin-down, DM ($\mathrm{d}$)\dotfill &  57494.619010\\ 
\hline
\multicolumn{2}{c}{Derived Quantities} \\ 
\hline
ECC, Eccentricity\dotfill &  0.000003 \\ 
OM, Longitude of periastron ($\mathrm{{}^{\circ}}$)\dotfill &  287.358425\\ 
\hline
\multicolumn{2}{c}{\footnotesize{a: Values quoted in Section~\ref{sec:pintmcmc} should supersede this $\chi^2$ fit.}}
\end{tabular}}
\end{table}

\subsection{\texttt{PINT} $\chi^2$ Gridding}\label{sec:grid}

The first set of analyses comprise \texttt{PINT}-based $\chi^2$ grids over various combinations of Shapiro delay parameters using all available GBT and Nan\c{c}ay data. This gridding routine is nearly identical to \citeauthor{fon16}~\citeyear{fon16}, which was based on \citeauthor{spl02} \citeyear{spl02}; however, the \cite{fon16} implementation uses \textsc{Tempo2}. The version used in this work performs a $\chi^2$ minimization using \texttt{PINT} for given combinations of companion mass and inclination angle on a 200$\times$200 grid, holding cos($i$) and \mc{} fixed and fitting for all other timing model parameters. As in \citeauthor{fon16}~\citeyear{fon16}, the $\chi^2$ map is transformed into a two-dimensional PDF and integrated along both axes to obtain one-dimensional PDFs for both Shapiro delay parameters. 

\begin{figure}
    \centering
    \includegraphics[width=0.9\textwidth]{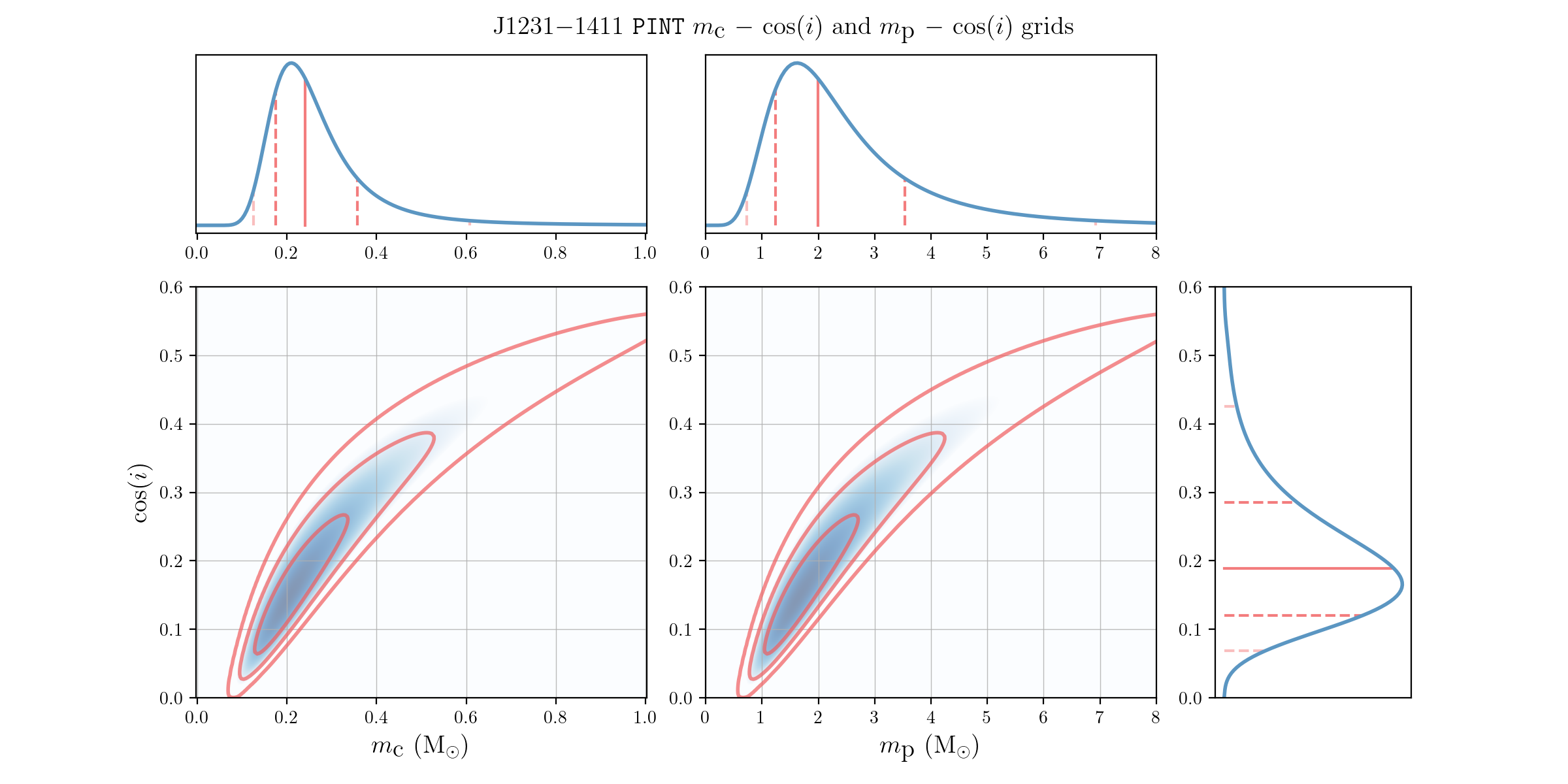}
    \caption{$\chi^2$ Gridding Results: The blue heat map shows the results of a $\chi^2$ minimization routine over cos($i$) and $m_{\mathrm c}$. Darker blue areas indicate lower values of $\chi^2$. Red contours denote 1, 2, and 3-$\sigma$ significance regions. Probability density functions denoted by blue lines on the top and side of the plot are projected from the $\chi^2$ maps, with dashed red lines denoting the 1 and 2-$\sigma$ confidence intervals (solid center lines are the distribution median).}
    \label{fig:1231grid}
\end{figure}

Figure~\ref{fig:1231grid} shows the resulting $\chi^2$ map and projected probability density functions for a trial that includes the white noise parameters in Table~\ref{tab:WN} (held fixed), parallax (PX), proper motion in RA and DEC (PMRA, PMDEC), an isotropic solar wind model that falls as $r^{-2}$ (SWM0 as implemented in \texttt{PINT}), and $\dot{P_{\rm b}}$ as well as a Taylor-expansion model of the dispersion measure (DM) to two terms (DM1, DM2). With a minimum elongation of 9 degrees, delays from propagation in the solar wind are negligible. Employing alternate DM models (e.g.~a single DM parameter or epoch-dependent DMX parameters) did not have an appreciable impact on the measured $m_{\textnormal c}$ or cos($i$) as seen in~\cite{fon21}; this is expected given the breadth of observing frequencies and high cadence of that data set. The inclusion of power-law red noise is not favored when these timing model parameters are present. From this trial, we find modest constraints on companion mass, inclination angle, and pulsar mass, with \mc{} = $0.24^{+0.12}_{-0.07}$ \msun{}, inclination $i$ = $79.11^{+3.96}_{-5.69}$\degree, and \mp{} = $2.00^{+1.53}_{-0.76}$ \msun{} (68.3\% CI). Due to its faster speed, this method was effective for more thoroughly exploring alternate timing and noise models. Although it is possible to apply a prior to parameters in the $\chi^2$ gridding scheme, covariances between, e.g.~binary and astrometric parameters, as well as the poorly-constrained companion mass, make the fully Bayesian approach described in Section~\ref{sec:pintmcmc} a more suitable choice for PSR \psr{}.

\begin{deluxetable*}{lccl}
\tablewidth{0.99\linewidth}
\tablecaption{\label{tab:WN}Summary of White Noise Parameters and JUMP values derived with \texttt{ENTERPRISE} and \texttt{PINT}}
\tablecolumns{4}
\tablehead{
\colhead{Backend \& Receiver} &
\colhead{EFAC}&
\colhead{EQUAD ($\mu$s)}&
\colhead{JUMP ($\mathrm{s}$)}
}
\startdata
GBT 350 MHz & 1.360 & 0.053 & $0.00155(3)$ \\
GBT 820 MHz (2009) & 1.745 & 0.079 &  $-5(6) \times 10^{-6}$\\
GBT 820 MHz (2019) & 1.146 & 3.838 &  $0.002222(2)$\\
NRT L-band, BON (2009) & 0.691 & 0.234 &  $3.6(6) \times 10^{-6}$\\
NRT L-band, NUPPI (pre-58600) & 1.061 & 0.073 & -- \\
NRT L-band, NUPPI (post-58600) &  1.064 & 0.124 &  $-1.1(5) \times 10^{-6}$\\
\enddata
\end{deluxetable*}

\subsection{\texttt{PINT} MCMC trials}\label{sec:pintmcmc}

Bayesian approaches to pulsar timing offer numerous benefits over traditional pulsar timing techniques for a variety of specialized applications \citep[e.g.][]{vig14}; however, the associated computational demand has prevented these methods from seeing widespread use. For example, the ability to include astrophysically motivated priors is a powerful improvement over typical least-squares techniques. This is of particular interest to astronomers wishing to conduct pulsar timing fits using complementary data sets (for example, one could use astrometric measurements from \textit{Gaia} to constrain a pulsar's parallax or proper motion). The \texttt{PINT} software package employs \texttt{emcee}, a widely-used open-source Python implementation of an affine-invariant ensemble sampler for MCMC \citep{for13}, to facilitate Bayesian timing model fits.

\subsubsection{Priors on $m_{\mathrm c}$ and \sini{}}\label{sec:priors}

The evolutionary end-point of a low-mass white dwarf binary (that is, the system's total mass and orbital period) can be directly predicted from the progenitor red giant star's degenerate core mass and radius, and is strongly influenced by the star's metallicity $Z$~\citep{ref71}. In their 1999 work, Tauris \& Savonije formulated a relationship between the binary orbital period of a pulsar system and the white dwarf companion's mass (TS99; \citeauthor{tau99} \citeyear{tau99}, see also \citeauthor{ist14} \citeyear{ist14} and \citeyear{ist16}):
\begin{equation}
\frac{M_{\mathrm WD}}{M_{\odot}} = \left(\frac{P_{\mathrm b}}{b}\right)^{(1/a)} + c.
\end{equation}
Values of $a$, $b$, and $c$ are dependent on the chemical composition of the companion: Pop.~I stars (young and metal-rich; $Z$ = 0.02) have ($a$ = 4.50, $b$ = $1.2\times 10^5$, $c$ = 0.120); the median Pop.~I/II (medium-$Z$) values are (4.75, $1.1\times 10^5$, 0.115); and the Pop.~II (old, metal-poor; $Z$ = 0.001) values are (5.00, $1.0 \times 10^5$, 0.110). This relationship holds for companions with masses between 0.18 and 0.45 \msun{}. A recent effort \citep{mat20} extrapolates this relationship to lighter white dwarfs using the relationships outlined in \cite{ist16}. For \psr{}'s 1.86-day orbit, the TS99 Pop.~I, Pop.~I/II, and Pop.~II values are 0.205, 0.214, and 0.223 \msun{}, respectively. Priors on $m_{\mathrm c}$ were chosen based on TS99 predictions (see Section~\ref{sec:MCMCresults}).

The most agnostic prior on inclination angle is that which describes a random distribution of inclinations; that is, a ``flat in cos($i$)'' distribution. 
Pulsar timing software typically uses the parameter sin($i$), for which the prior becomes
\begin{equation}
p(z) = \frac{z}{\left(1-z^2\right)^{1/2}}
\end{equation}
for values of $z\equiv \mathrm{sin}(i)$ between 0 and 1.

The data are informative themselves, and do not ``require'' the \mc{} or \sini{} priors for a significant measurement. However, a flat prior on \mc{} permits large, unphysical swaths of parameter space, and omitting the prior on \sini{} introduces an inaccurate bias in expected orbital orientations.
In placing only modest priors on \mc{}, we have elected to take a more conservative approach in an effort to understand the underlying physical processes rather than to obtain an improved pulsar mass constraint. However, in a context such as the X-ray lightcurve analysis of \cite{sal24}, a restriction of parameter space to physical values is warranted (see Section~\ref{sec:discussion}). 

\subsubsection{Results of the Fixed-Noise MCMC Fits}\label{sec:MCMCresults}

We present the results from several MCMC trials using the radio data set. Each trial used 256 walkers and 3000 steps, with convergence ensured by calculating the autocorrelation time ($\tau$) for each chain in the fit \citep[see][]{goo10}. Suggestions for chain lengths vary, though \texttt{emcee} recommends values $>50\tau$. For each run, we have ensured that our integrated autocorrelation times and chain lengths meet that criterion, though many runs have factors of 1000 or more.

\begin{figure}
    \centering
    \includegraphics[width=0.55\textwidth]{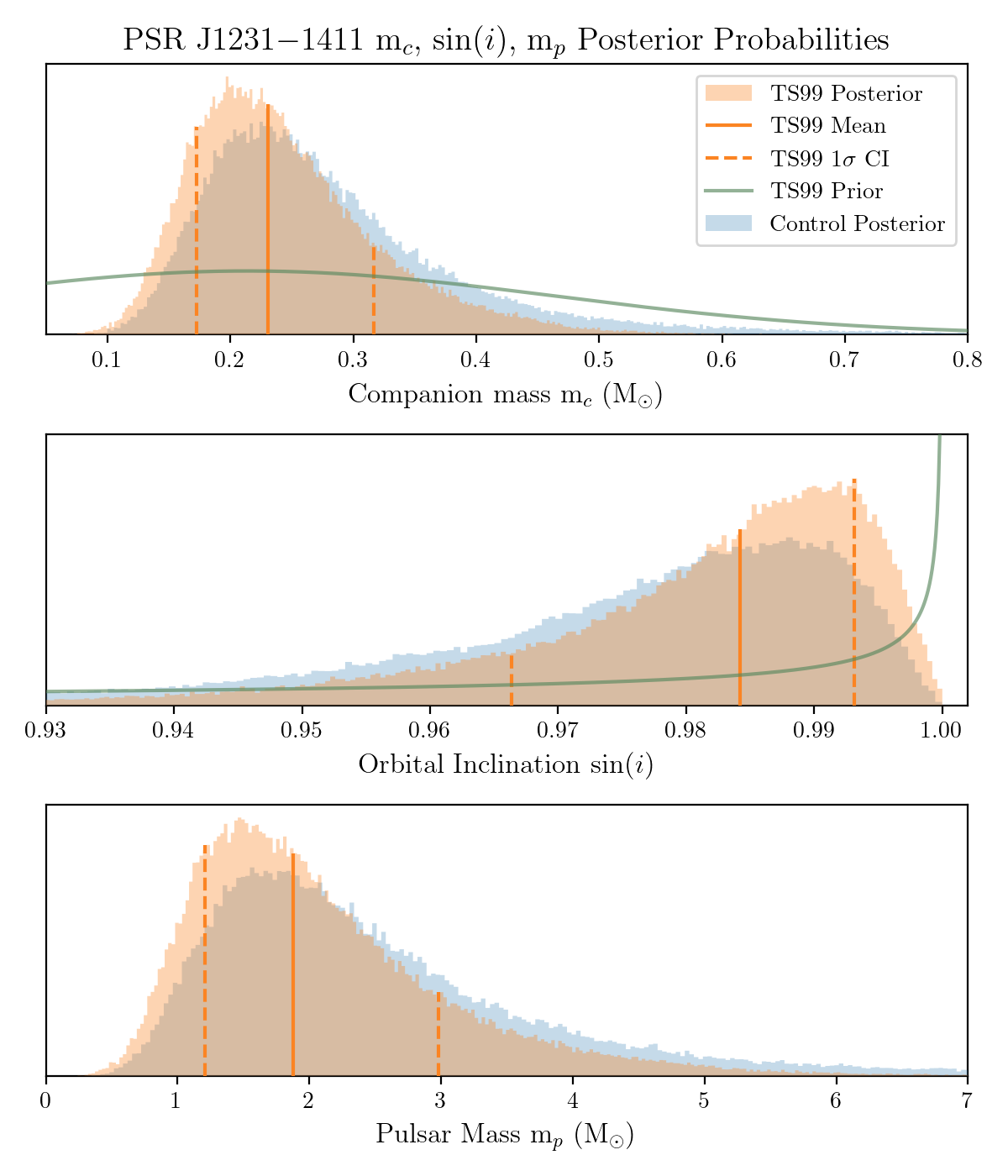}
    \caption{Top to bottom, posterior distributions for \mc{}, \sini{}, and \mp{} for two MCMC trials. White noise parameters have been determined \emph{a priori} and are held fixed; all other parameters are free. The orange histogram is the PDF for a trial using both priors (green lines) described in Section~\ref{sec:priors}. The 68.3\% CI is shown with orange lines. The blue PDF behind the TS99 distribution shows the results of a fit with flat priors applied.}
    \label{fig:1231MCMC}
\end{figure}

Figure~\ref{fig:1231MCMC} shows PDFs for \mc{}, \sini{}, and the derived \mp{} for two runs. The first (in orange), referred to as ``TS99,'' employs a wide Gaussian prior ($\sigma$ = 0.25\,M$_{\odot}$, bounded between 0 and 1\,M$_{\odot}$) on \mc{} centered around the predicted intermediate-metallicity (Pop.~I/II) companion mass value of 0.214\,\msun{} as well as the ``flat in cos($i$)'' prior. The second, designated ``control,'' uses flat priors for both parameters---a deliberately poor choice in the case of \sini{}. In assessing the \mc{} prior, we also checked the impact of constraining the companion's mass to a more physical maximum of 0.5\,\msun{}; however, the impact on \mp{} is minimal ($<2\%$) and the result is not plotted. The TS99 fit yields \mc{} = $0.23^{+0.09}_{-0.06}$ \msun{}, inclination $i$ = $79.80^{+3.47}_{-4.70}$\degree, and \mp{} = $1.87^{+1.11}_{-0.67}$ \msun{} (68.3\% CI). Imposing an additional prior disallowing pulsar mass values greater than 3 M$_{\odot}$ yields \mc{} = $0.22^{+0.06}_{-0.05}$ \msun{}, inclination $i$ = $80.55^{+3.03}_{-3.21}$\degree, and \mp{} = $1.71^{+0.70}_{-0.56}$ \msun{} (see Figure~\ref{fig:nonlinPDF}). The ``control'' (flat-prior) fit results in slightly worse constraints, with \mc{} = $0.26^{+0.11}_{-0.07}$ \msun{}, inclination $i$ = $78.09^{+3.99}_{-5.46}$\degree, and \mp{} = $2.21^{+1.55}_{-0.82}$ \msun{}.

\subsection{Results of the Fully-Bayesian Nonlinear MCMC Fits}\label{sec:nonlinearresults}

One drawback of the fits discussed thus far is the fact that noise models are determined \emph{a priori}, which might hide covariances between timing and noise model parameters. In order to simultaneously sample both noise and timing models, we instead use \texttt{ENTERPRISE}, rather than a serial combination of \texttt{ENTERPRISE} and \texttt{PINT}'s MCMC sampling routine, to determine the best-fit solution. 

Under normal circumstances, \texttt{ENTERPRISE} assumes that timing residuals represent linear deviations from the best-fit parameter values, constructing a design matrix where offsets are limited to uniform deviations around those values. Analytical marginalization over those deviations can hide nonlinear responses that may be relevant for some timing model parameters. 
To address these concerns, Kaiser et al.~(2025, in prep.) are developing a branch of \texttt{ENTERPRISE} that allows for a fully generalized Bayesian approach where the timing model is incorporated into the MCMC routine as a deterministic signal. One is also able to designate some parameters as nonlinear while allowing others to undergo analytical marginalization in the typical way. For parameters specified as nonlinear, a penalty is imposed through the addition of an MCMC step that recomputes residuals and incorporates them into the \texttt{ENTERPRISE} likelihood calculation, allowing for a full exploration of the parameter space. Early results from their work, which applies this new scheme to several binaries in the NANOGrav 15-year data set, indicates that these changes may result in improved mass constraints for certain MSP binaries.

As a system with measurable but poorly constrained Shapiro delay, PSR \psr{} presented an opportunity to ensure no significant covariances between white noise parameters and timing model parameters exist, as well as glimpse what improvements might be possible with more sophisticated analysis. As in the \texttt{PINT} MCMC trials, we used the data set with the same JUMPs and timing model parameters. Pulsar mass values are constrained to a maximum of 3\,\msun{}, a significant difference from our previous trials. A prior on parallax determined by the pulsar's DM distance (as in \citeauthor{vig14} \citeyear{vig14}, using the \texttt{NE2001} electron density distribution from \citeauthor{cor02} \citeyear{cor02}) is applied, taking the form:
\begin{equation}
    p(\textnormal{PX}) = \frac{1}{\sqrt{2\pi}\sigma_d\textnormal{PX}^2}\textnormal{exp}\left[-\frac{(\textnormal{PX}^{-1}-d)^2}{2\sigma_d^2}\right]
\end{equation}
where PX > 0 and the DM distance $d$ is Gaussian with variance $\sigma^2_d$. 
We find that all well-measured timing parameters agree closely with those measured in the previous gridding and MCMC trials, with the exception of PX, which is lower by 2$\sigma$ and slightly better constrained, likely due to the imposed prior.

\begin{figure}
    \centering
    \includegraphics[width=0.6\textwidth]{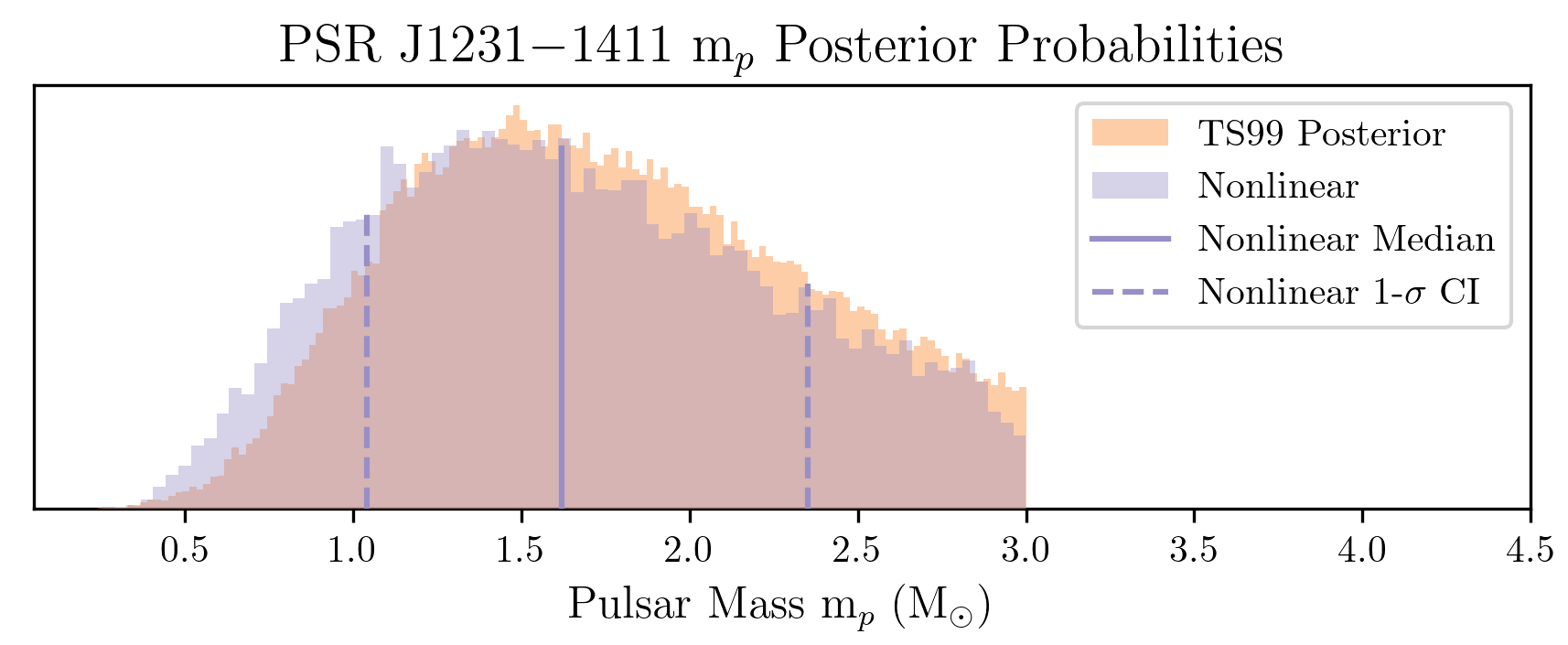}
    \caption{The posterior probability distribution for \mc{} from the fully-generalized fit is shown in purple, with the median and 1-$\sigma$ CI overplotted. The orange histogram results from the fixed-noise TS99 fit shown in Figure~\ref{fig:1231MCMC} with the additional 3 M$_{\odot}$ prior applied.}
    \label{fig:nonlinPDF}
\end{figure}

As before, we find no significant red noise. 
Posterior distributions for all noise parameters measured in the \emph{a priori} linear analysis and the generalized nonlinear analysis closely agree, with all median EQUAD and EFAC values varying by less than 5\%. 
From this exploratory analysis, we find the companion mass \mc{} = $0.21^{+0.06}_{-0.05}$ \msun{}, inclination $i$ = $81.06^{+3.36}_{-3.51}$\degree, and \mp{} = $1.62^{+0.73}_{-0.58}$ \msun{} (68.3\% CI). The derived mass and inclination values from the nonlinear runs agree with those from our MCMC trials, which is expected given the large uncertainties. However, we find that the new approach yields improved constraints on the orbital inclination and companion mass (see Figure~\ref{fig:nonlinPDF}). An observed decrease in \mp{} is largely attributable to the imposition of a 3\,\msun{} pulsar mass constraint, which is reflected in the close agreement seen between the fully generalized measurements and the fixed-noise MCMC results when this maximum mass prior is applied. Taken together, these results suggest that in the case of PSR \psr{}, determining the white and red noise values prior to MCMC fits is a reasonable approach (see Section~\ref{sec:discussion}). The improvements brought by this nonlinear technique merit further exploration, though the associated computational demand is high (see the upcoming paper by Kaiser et al.). 

\section{Discussion}\label{sec:discussion}

We explored a large number of model permutations during our analysis, such as the inclusion of poorly measured astrometric parameters, different approaches to DM and solar wind timeseries modeling, and different Shapiro delay parameterizations. We applied each of these in turn to the whole data set as well as its constituent backend-receiver-telescope members as individual data sets, though only results using the full data set are presented here (with the exception of several illustrative comparisons presented later in this section). While a detailed description of each of these trials is beyond the scope of this work, we found many to have a significant impact on measured white and intrinsic red noise, and in turn the pulsar and companion masses. We note that \psr{} is a somewhat unusual case in the landscape of precision pulsar mass measurements. In the ``gold-standard'' examples of Shapiro delay-derived masses, the systems (many of which are observed with pulsar timing arrays, and are therefore selected against strong scintillation or orbital variability) are sufficiently edge-on that the Shapiro delay effect is significantly more pronounced, rendering the mass measurement more independent of the analysis methods and assumed model.

Because of PSR \psr{}'s proximity and large proper motion, measurement of long-term secular orbital effects such as $\dot{P_{\rm b}}$ (change in binary orbital period), $\dot{x}$ (change in projected semi-major axis), and $\dot{\omega}$ (change in longitude of periastron) should be considered. For the first time to our knowledge, we report a significant $\dot{P_{\rm b}}$ and parallax measurement for \psr{}. Observed $\dot{P_{\rm b}}$ is not solely (or even notably, in the case of PSR \psr{}) attributable to a loss of orbital energy through gravitational radiation. Galactic differential rotation, z-axis acceleration within the Galactic potential, and the Shklovskii effect---a seeming acceleration due to astrometric parallax---all contribute to the total observed orbital period decay. All four of these effects are fundamentally defined by the individual component masses and distance to the system in addition to the well-measured timing parameters; therefore, an independent measurement of \mc{} and \mp{} with two significantly detected post-Keplerian parameters can be compared for consistency with a significantly measured parallax, as we see with PSR \psr{}. Section 4.2 of \citeauthor{fon21}~\citeyear{fon21} (and references therein) presents a thorough overview of the methods and quantities used here, including explicit definitions for each component of $\dot{P_{\rm b}}$. We have chosen to use the same Galactic speed and distance parameters cited in that work \citep{gra19} as well as the traditional model of Galactic gravitational potential \citep{kui89}. We similarly do not include any negligible higher-order corrections to $\dot{P_{\rm b}}$ \citep{hu20}.

\begin{figure}
    \centering
    \includegraphics[width=0.7\textwidth]{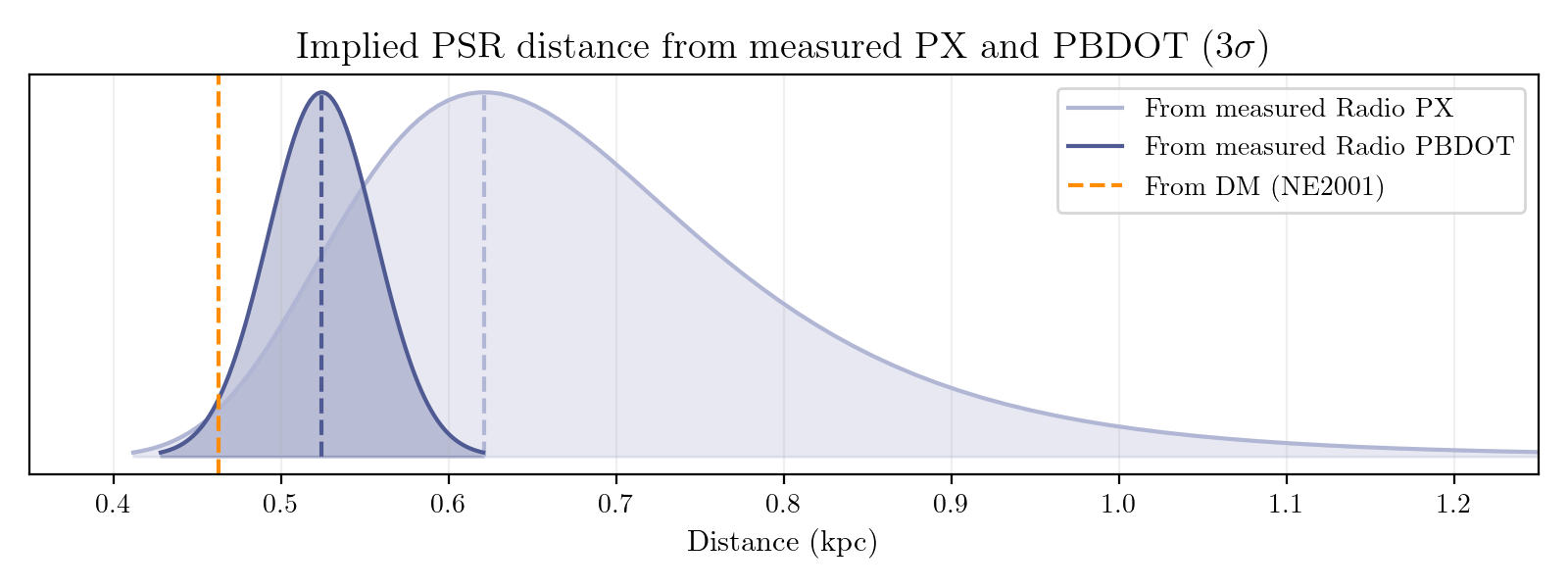}
    \caption{Derived pulsar distance measurements for PSR \psr{} from parallax and $\dot{P_{\rm b}}$ in the radio analysis, normalized to equal heights for illustration purposes only. The lighter PDF is the derived distance from measured timing parallax, while the darker PDF is from $\dot{P_{\rm b}}$ constraints. The orange dashed line represents the predicted value from the NE2001 elctron density model \citep{cor02}.}
    \label{fig:pbdot}
\end{figure}

Figure~\ref{fig:pbdot} summarizes the distance to PSR \psr{} as derived from radio measurements of $\dot{P_{\rm b}}$ and PX in addition to other relevant timing parameters. The DM-derived distance of 0.463 kpc from the NE2001 electron density model (as implemented in \texttt{NE2001p} by \citeauthor{ock24} \citeyear{ock24}) is also plotted for reference. Though $\dot{P_{\rm b}}$ is significantly better constrained than parallax, the two derived distance values are in agreement. At the time of the \cite{sal24} analysis, our radio timing model had erroneously excluded $\dot{P_{\rm b}}$, as a recent addition of data caused the parameter to gain significance. Although the PX-derived distance constraint of $\sim$600$\pm$100 pc was provided, a more conservative (uniform) prior between 100 and 700 pc was applied in the X-ray lightcurve modeling.

As PSR \psr{}'s orbit is nearly circular, secular variations in $\dot{P_{\rm b}}$, $\dot{x}$, or $\dot{\omega}$ due to intrinsic changes are not expected. However, just as an apparent $\dot{P_{\rm b}}$ was significantly measured, so too might the related non-relativistic (apparent) $\dot{x}$ and $\dot{\omega}$. As in \cite{ala21}, an F-test with p-value threshold of 0.0027 was conducted to determine whether the inclusion of these post-Keplerian parameters was appropriate. While $\dot{P_{\rm b}}$ was \emph{strongly} inconsistent with noise, $\dot{x}$ is borderline ($\sim$2$\sigma$) and $\dot{\omega}$ is not at all significant; therefore, the latter two parameters were not included in the model. The maximum contribution to $\dot{x}$ from proper motion is given by 
\begin{equation}
\left(\frac{\dot{x}}{x}\right)^{\rm PM} \leq 1.54 \times 10^{-16} {\rm cot}\,i \left(\frac{\mu_{\rm T}}{{\rm mas}\,{{\rm yr}^{-1}}}\right),
\end{equation}
where $\mu_{\rm T}\equiv\sqrt{\textnormal{PMRA}^2 + \textnormal{PMDEC}^2}$ (the sum of the squares of proper motion in RA and DEC; see \citeauthor{lor04} \citeyear{lor04}). Our best-fit values of $i$ and $\mu_{\rm T}$ yield a maximum $\dot{x}$ due to proper motion of 2.04$\times 10^{-14}$. As a point of reference, the current constraint on $\dot{x}$ predicts a maximum inclination of $i\leq85\degree$ (1$\sigma$ upper limit). With additional data, especially from next-generation facilities where timing precision will improve, $\dot{x}$ may be significantly constrained.

Continued Nan\c{c}ay timing may improve constraints on \mp{}; however, returns diminish as orbital phase becomes well-sampled. As an illustration, between MJD $\sim$56100 (the point at which Shapiro delay becomes measurable) and 56750---the first 1.8 years of observations---uncertainty in \mp{} decreased by 1\,M$_{\odot}$. Between MJD 59000 and the data set's last observation on 60408 (the last $\sim$3.8 years of data), uncertainty decreased by only 0.15\,M$_{\odot}$. Since MJD 59750, the most recent 1.8 years of data, uncertainty on \mp{} is unchanged. An additional avenue for future exploration may be the use of PSR \psr{}'s strong scintillation to obtain a series of scintillation arc measurements, which could yield constraints on the transverse velocity and in turn, independently measure the system's inclination and longitude of ascending node \citep{cor98}.

Observing relatively dim MSPs like PSR \psr{} at L-band and potentially higher frequencies with high-gain, next-generation radio telescopes such as the DSA-2000, ngVLA, and SKA will improve the associated timing precision with less chromatic noise, thus improving measured \mp{}. An analysis incorporating available MeerKAT data by the RelBin team is underway for PSR \psr{}. Scheduling observations at specific orbital phases can be challenging as-is, but if a continuous campaign over a single epoch were possible with a next-generation facility, any uncertainty due to changing DMs could be minimized. As advanced radio telescopes come online, both the number of new pulsar discoveries as well as higher achievable timing precisions will facilitate improved mass measurements. 

\begin{figure}
\centering
\includegraphics[width=0.8\textwidth]{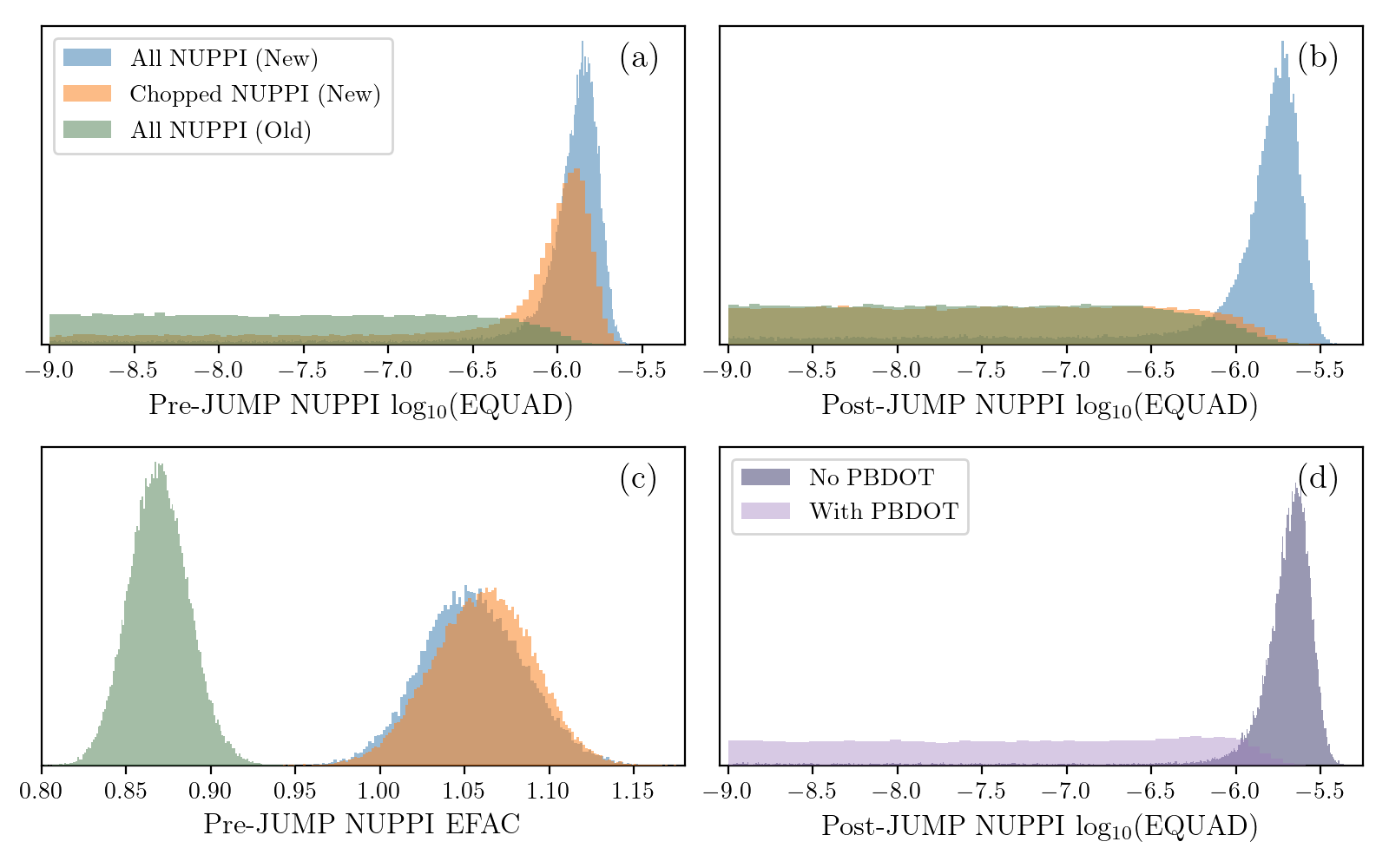}
\caption{Sensitivity of measured EFAC and EQUAD to a lengthened data set, improved NRT polarization calibration, and $\dot{P_{\rm b}}$ inclusion. Trials are conducted with NUPPI data only. (a) Compared to the new, longer NUPPI data set (blue) with measured \mp{} = 2.64$^{+3.88}_{-1.33}$ \msun{}, the old NUPPI data (green) yielded better constraints with \mp{} = 0.99$^{+0.86}_{-0.43}$ \msun{}, which was not expected when using a shorter-timespan data set (see Figure~\ref{fig:slices}). To control for differences between the old and new TOA generation methods, the new NUPPI data were chopped to the length of the old NUPPI data (orange). The constraint was in-between in both measured mass and uncertainty (\mp{} = 1.31$^{+1.12}_{-0.62}$ \msun{}), and did not worsen to the extent expected with such a decrease in timing baseline. The \mp{} constraint worsened with the full data set because these new data were sufficiently improved to require $\dot{P_{\rm b}}$, which is not included in any of the fits in panel (a) and is being compensated for by EQUAD. When PBDOT is added to the model, the new EQUAD value closely follows the old (green) distribution, and constraints on \mp{} are improved. (b) On the other hand, when new NUPPI data in the post-JUMP era are chopped to the length of the old data set, they do not constrain EQUAD, because the length of the data set after the MJD 58600 JUMP is much shorter. (c) Old, improperly calibrated NUPPI data (green) likely had overestimated uncertainties leading to EFAC < 1. (d) In a fit to the full-length, new data set with (light purple) and without (dark purple) PBDOT, the lack of PBDOT manifests as a large, well-constrained EQUAD.}
    \label{fig:noise}
\end{figure}

Because we undertook the timing analysis in stages, adding NUPPI data as it became available, several interesting findings emerged. The initial (``old'') data set included NUPPI data through MJD 59754, clock corrections contaminated by the NRT GPS error, and less precise polarization calibration. The ``new'' NUPPI data set used in this analysis included updated data, proper clock corrections, and improved polarization calibration, the latter resulting in improved TOA uncertainties ($\sim$10\,$\mu$s vs.~6.7\,$\mu$s). The old data likely had overestimated error bars as evidenced by the measured EFAC value being less than one (see Figure~\ref{fig:noise}, panel c), which remains true for BON TOAs. 

In early models using the old data set, intrinsic red noise with a flat spectral index (i.e.~white noise) and high amplitude was found to be significant, but the spectral index increased to more typical values when updated NRT clock corrections were applied to an otherwise-identical set of TOAs. Because the inclusion of intrinsic red noise significantly worsens the constraint on \mp{} (increasing the positive 1-$\sigma$ uncertainty by factors of >5 in some cases), it is worth emphasizing its impact for pulsars with borderline-significant red noise that may be affected by small changes to the timing model. 

In earlier versions of this analysis, we derived fixed white noise parameter values from \texttt{ENTERPRISE} chains using the maximum likelihood value (MLV); the preferable method, presently utilized, is to average the top N highest likelihood values (the mean of large likelihoods, MLL) to obtain a more stable estimate (here, N=50). We conducted trials to determine if these two methods were discrepant enough for PSR \psr{} to cause significant differences in measured timing parameter values, finding that differences were negligible. 

\begin{figure}
    \centering
    \includegraphics[width=0.5\textwidth]{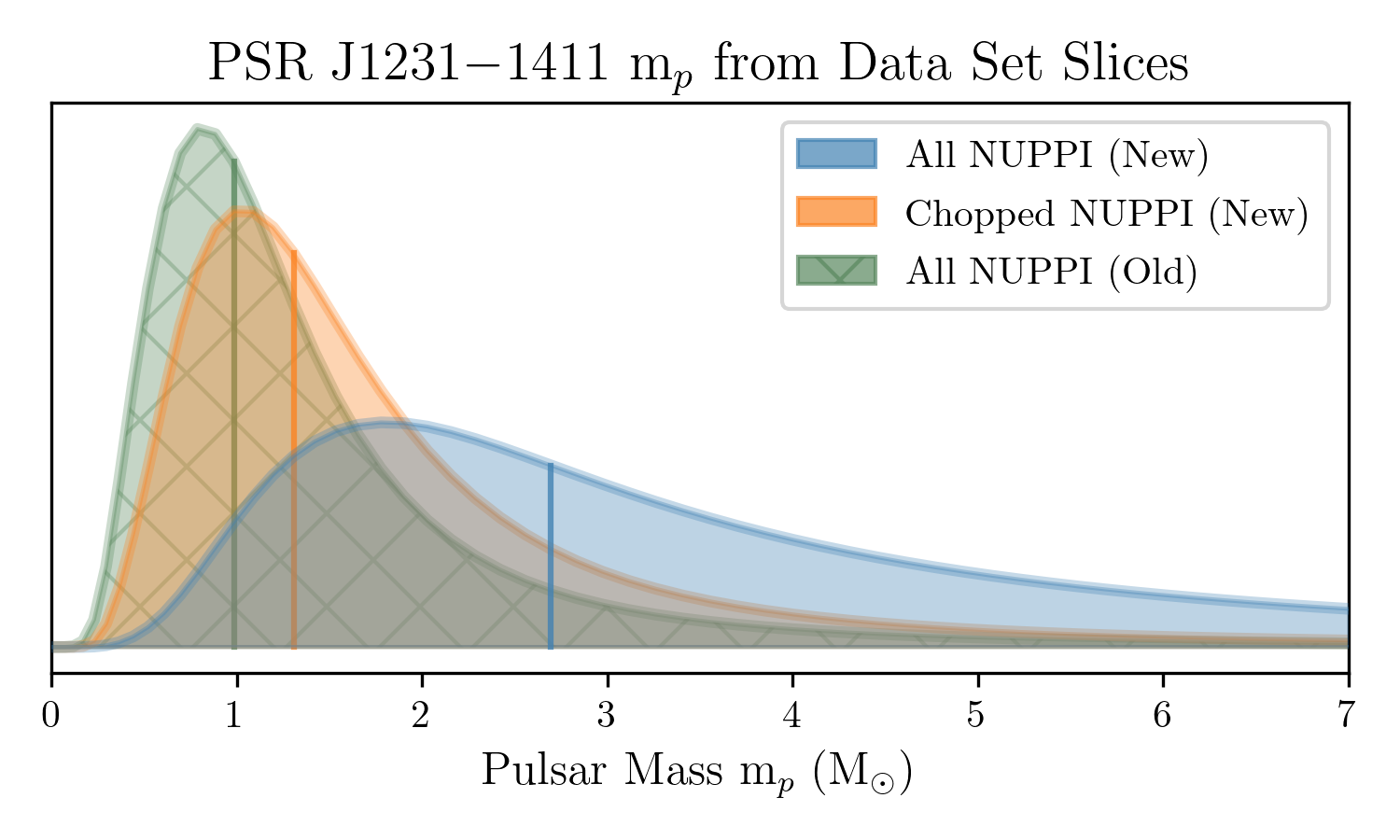}
    \caption{Posterior PDFs for \mp{} as measured for the three trials presented in Figure~\ref{fig:noise}, none of which include $\dot{P_{\rm b}}$. Colors are consistent between the two figures. Because it requires a significant, well-measured EQUAD, the new NUPPI data set yielded a significantly worse \mp{} than the older data set (2.64$^{+3.88}_{-1.33}$ \msun{} and 0.99$^{+0.86}_{-0.43}$ \msun{}, respectively). }
    \label{fig:slices}
\end{figure}

While one would expect constraints on \mp{} to improve with the addition of more---and better---data, we found that the uncertainty in \mp{} increased when new NUPPI data were initially added. Trials were then conducted with a limited, NUPPI-only data set to test this behavior. When new NUPPI data were cut to the length of the old data set in an effort to control for the improved clock corrections and TOA uncertainties, constraints once again improved, suggesting that the shorter data set was doing a ``better'' job despite more data being available (see Figure~\ref{fig:noise} caption for details). The discrepancy lies with EQUAD, which was well-constrained to a relatively high value in the new, full NUPPI data set (because $\dot{P_{\rm b}}$ had not yet been added to the model), but poorly constrained in the shorter old and chopped data sets (see Figure~\ref{fig:noise} panels a, b). In comparing the measured EQUAD before and after the addition of $\dot{P_{\rm b}}$ (Figure~\ref{fig:noise}), we find that the well-constrained, high EQUAD leads to significantly poorer mass and inclination constraints. As a point of reference, the pulsar mass and inclination derived for the old NUPPI data set (i.e.~with low EQUAD) during one of these trials were 0.99$^{+0.86}_{-0.43}$ \msun{} and $83.60^{+3.49}_{-5.88}$\degree, while the values for the new NUPPI data were 2.64$^{+3.88}_{-1.33}$ \msun{} and $75.34^{+6.51}_{-9.23}$\degree (see Figure~\ref{fig:slices}). The position of the JUMP, which was originally placed at MJD 58077 (corresponding to an issue with the NRT's focal carriage) instead of 58600 (when a pre-amplifier was replaced), did not affect measured white noise parameter values, and are identical to the green ``All NUPPI (Old)'' posteriors in Figure~\ref{fig:noise}. Despite this, the original NUPPI TOAs show significant red noise when fit with the more recent MJD 58600 JUMP, and therefore a worse mass and inclination constraint comparable to the ``Chopped NUPPI (New)'' data. The sensitivity of \mp{} to both white and red noise values should not be ignored for pulsars with relatively poor timing precision. 

Notably, EQUAD was well-constrained in the 2019 GBT campaign both with and without $\dot{P_{\rm b}}$. This indicates that these data, which come from lower-frequency observations than data from NRT, show more chromatic effects from scattering and DM variations. Because the data set is relatively short and more sophisticated chromatic modeling is not required for the NRT data, the larger EQUAD for the 2019 GBT data is acceptable.

We find that the simultaneous determination of noise and timing model parameters described in Section~\ref{sec:nonlinearresults} is better able to explore the parameter space when $\dot{P_{\rm b}}$ is not included and EQUAD is therefore high, yielding quite similar constraints on the Shapiro delay parameters regardless of the inclusion of $\dot{P_{\rm b}}$. When noise is instead determined \emph{a priori} and fixed during the MCMC sampling, the same flexibility is not observed; presumably because all MCMC samples have the same high EQUAD, the \mp{} constraint is notably worse in the fits without $\dot{P_{\rm b}}$. 

We encourage the reader to consider these cautionary tales relevant only to binaries with poor timing precision, lower inclinations (than e.g.~80-85 degrees), or other detrimental features that result in unconstrained \mp{}. For well-measured, near-edge-on binaries like PSR J1614$-$2230 or J1909$-$3744, the relativistic Shapiro delay remains uniquely poised to constrain neutron star masses with little dependence on modeling choices. 

The measured PSR \psr{} companion mass \mc{} = $0.23^{+0.09}_{-0.06}$ \msun{} is consistent with the TS99 prediction both with and without TS99 priors applied, but is not well-constrained enough to glimpse the system's detailed evolutionary history. Our findings are also in agreement with analyses of optical photometric and spectroscopic observations of PSR \psr{}'s companion using the Gemini South telescope as presented in \citeauthor{tes15} \citeyear{tes15} and \citeauthor{bas16} \citeyear{bas16}. The former discovered the optical counterpart, finding that it must be cool and old ($T_{\textnormal{eff}} \leq 3000$ K and age $\geq 7$ Gyr), with a distance $d \leq 0.5$ kpc. \citeauthor{bas16} \citeyear{bas16} reported that companion masses between 0.188--0.384 M$_{\odot}$ and distances $d \sim$ 0.35--0.51 kpc  are the allowable ranges to reproduce the optical and radio results using their chosen helium white dwarf models. If the companion is confirmed to have $T_{\textnormal{eff}} < 3000$ K, it could be one of the coolest white dwarfs known. The optical results, taken together with our long-term timing effort, have indicated that with a mildly-inclined orbit of $79.80^{+3.47}_{-4.70}$\degree and inferred \mp{} = $1.87^{+1.11}_{-0.67}$ \msun{} (or $1.62^{+0.73}_{-0.58}$ \msun{} from the simultaneous-sampling trials presented in Section~\ref{sec:nonlinearresults}), PSR \psr{} is likely a typical MSP-white dwarf binary.

Despite large uncertainties on \mc{} and \mp{}, these results---especially the relatively strong inclination constraint---still informed the NICER X-ray profile analysis \citep{sal24}. The mass-inclination priors employed in that work were based on a preliminary result using tighter \mc{} priors in our MCMC fit (yielding \mp{} = $1.48^{+0.81}_{-0.53}$ and inclination $i$ = $80.98^{+3.19}_{-4.11}$\degree), and further modified by imposing a strict 1--3 \msun{} limit on the distribution, resampling, and applying further physical priors based on neutron star compactness. The resulting \mp{} prior was essentially uniform over the chosen mass range. Given the faint Shapiro delay and troublesome radio timing for PSR \psr{}, a broad mass prior is expected. However, our $\sim$5\% constraint on orbital inclination, which is significantly more resilient to changes in fitting method, red or white noise inclusion, and chosen priors, could be considered the more informative addition to the analysis.

Posterior distributions for \mp{}, \mc{}, sin($i$), and PX for the MCMC trials presented in Section~\ref{sec:pintmcmc} and Section~\ref{sec:MCMCresults} are available on Zenodo: \url{10.5281/zenodo.17596801}. 

\section{Conclusions}\label{sec:conclusion}

We have conducted a radio timing analysis for the $\gamma$-ray-bright MSP \psr{} in order to constrain Shapiro delay and in turn, the pulsar's mass. This study introduces 22 hours of new GBT observations for the source in combination with 15 years of archival NRT data and several epochs of archival GBT data. MCMC-enabled trials with \texttt{PINT} and a white noise model measured \emph{a priori} were informed by the TS99 white dwarf mass vs.~orbital period relationship, yielding \mc{} = $0.23^{+0.09}_{-0.06}$ \msun{}, inclination $i$ = $79.80^{+3.47}_{-4.70}$\degree, and \mp{} = $1.87^{+1.11}_{-0.67}$ \msun{} (68.3\% CI). For simultaneous MCMC sampling of the noise and timing models we find \mc{} = $0.21^{+0.06}_{-0.05}$ \msun{}, inclination $i$ = $81.06^{+3.36}_{-3.51}$\degree, and \mp{} = $1.62^{+0.73}_{-0.58}$ \msun{} (68.3\% CI). Simultaneously fitting for noise along with the timing model via a nonlinear MCMC approach with strict upper limits on \mp{} yielded a lower median mass and smaller uncertainties, though entirely consistent with measurements using the other methods. These results are closely consistent with TS99-predicted values of $m_{\mathrm c}$, even for the trials with less informative priors. 
For pulsars such as PSR \psr{} with not-quite-edge-on orbits and low-mass companions, the measurable Shapiro delay is quite subtle and necessitates high-precision timing. With relatively poor timing precision, detectable astrometric variations, and borderline-significant intrinsic red noise, subtle changes in the timing model were shown to affect the measured \mp{}; we urge caution in analyzing such systems, especially in cases where the results are used as priors in downstream analyses or where the measured mass is large enough to be of relevance to equation of state constraints. Next-generation facilities, especially when flexible scheduling is available, will improve attainable \mp{} constraints.

\acknowledgements
H.~T.~C. acknowledges funding from the U.S. Naval Research Laboratory (NRL). Portions of this work performed at NRL were supported by NASA and ONR. SMR is a CIFAR Fellow and is supported by the NSF Physics Frontiers Center award 2020265.

The National Radio Astronomy Observatory and the Green Bank Observatory are facilities of the National Science Foundation operated under cooperative agreement by Associated Universities, Inc.

The Nan\c{c}ay Radio Observatory is operated by the Paris Observatory, associated with the French Centre National de la Recherche Scientifique (CNRS). We acknowledge financial support from the ``Programme National de Cosmologie et Galaxies'' (PNCG) and ``Programme National Hautes Energies'' (PNHE) of CNRS/INSU, France.

\facilities{Green Bank Telescope (GUPPI), Nan\c{c}ay Radio Telescope (NUPPI)}

\bibliography{thebibliography}
\bibliographystyle{aasjournal}

\end{document}